\documentstyle[aps,preprint]{revtex}

\widetext
\draft
\tighten
\def\abstract{\if@twocolumn
\section*{Resumen}
\else \small
\begin{center}
{\bf Resumen\vspace{-.5em}\vspace{0pt}}
\end{center}
\quotation
\fi}

\def\thebibliography#1{\section*{Referencias\markboth
{REFERENCIAS}{REFERENCIAS}}
\list
{[\arabic{enumi}]}{\settowidth\labelwidth{[#1]}
\leftmargin\labelwidth
\advance\leftmargin\labelsep
\usecounter{enumi}}
\def\newblock{\hskip .11em plus .33em minus -.07em}
\sloppy
\sfcode`\.=1000\relax}

\begin{document}

\preprint{EFUAZ FT-96-26}

\title{Interacci\'on `Oscilador' de Part\'{\i}culas
Relativistas\thanks{Ciertas partes de este art\'{\i}culo
han sido presentados en el Seminario del IFUNAM, 19 de noviembre de 1993,
el Seminario de la EFUAZ, 25 de mayo de 1994 y en el Simposio de
Osciladores Arm\'onicos, Cocoyoc, M\'exico, 23-25 de marzo de 1994.
Enviado a ``Investigaci\'on Cientifica".}}

\author{{\bf Valeri V. Dvoeglazov}}

\address{
Escuela de F\'{\i}sica, Universidad Aut\'onoma de Zacatecas \\
Antonio Doval\'{\i} Jaime\, s/n, Zacatecas 98068, ZAC., M\'exico\\
Correo electronico:  VALERI@CANTERA.REDUAZ.MX}

\date{25 de junio de 1996}

\maketitle

\abstract{
Es una introducci\'on en el nivel accesible a las recientes ideas
en mec\'anica relativista de part\'{\i}culas con diferentes
espines, interacci\'on de cuales es del tipo oscilador. Esta
construcci\'on matem\'atica propuesta por M. Moshinsky pudiera proveer
aplicaciones en la descripci\'on de los processos mediados por los
campos tensoriales y en la teor\'{\i}a de los estados ligados.
}

\pacs{PACS: 12.90}

\newpage

\setlength{\baselineskip}{24pt}

\section{Introducci\'on}

Con el oscilador arm\'onico ha trabajado gran parte de su vida
el doctor Marcos Moshinsky~\cite{Mosh4}, alumno de Eugene Wigner y
el primer {\it Ph. D.} en  f\'{\i}sica de M\'exico, este tipo de
interacci\'on se ha sido aplcado a muchos problemas en F\'{\i}sica
Matem\'atica, F\'{\i}sica At\'omica y Molecular, \'Optica, F\'{\i}sica
Nuclear y Part\'{\i}culas Fundamentales. Desde 1992 se selebran
Simposios Internacionales de los problemas relacionados con el oscilador
\'armonico.  Otros  f\'{\i}sicos conocidos de M\'exico, Rusia y de los
EE.UU., tales como N. Atakishiyev (IIMAS, Cuernavaca y Baku, Azerbaijan),
L. C.  Biedenharn (Duke, EE. UU.), O.  Casta\~nos (ICN-UNAM), J. P.
Draayer (Louisiana), A.  Frank (ICN-UNAM), F. Iachello (Yale, EE. UU.), Y.
S.  Kim (Maryland, EE.  UU.), V.  I.  Man'ko (Lebedev, Mosc\'u) , M. M.
Nieto (LANL, EE. UU.), L. de la Pe\~na (IF-UNAM), Yu.  F.  Smirnov
(IF-UNAM y MSU, Mosc\'u), K.  B.  Wolf (CIC, Cuernavaca) trabajan en esta
\'area.  Entonces, el objetivo de dar a conocer una parte de estos
problemas a los estudiantes de la UAZ y otras instituciones de la
provincia mexicana tiene suficientes razones.

En esta Secci\'on me permito  echar una breve mirada al desarollo de estas
materias hasta los a\~nos noventas. Tanto en mec\'anica cl\'asica como en
mec\'anica cu\'antica no relativista los problemas de movimiento del
una part\'{\i}cula con masa en un punto  en el campo potencial
`oscilador' $V(x) = {1\over 2} Kx^2$ pueden ser resueltos en forma
exacta. La ecuaci\'on de Schr\"odinger de la mec\'anica cu\'antica
\begin{equation}
-{\hbar^2 \over 2m} \frac{d^2 \Psi (x)}{dx^2} +{1\over 2} Kx^2 \Psi (x)
=E\Psi (x)
\end{equation}
nos da valores propios de energia que son discretos ($\omega=\sqrt{K/m}$):
\begin{equation}
E_n = \hbar \omega (n+{1\over 2})\quad,\quad n=0,1,2\ldots
\end{equation}
Este tipo del espectro se llama espectro {\tt equidistante}.
En este problema tenemos la energ\'{\i}a {\tt zero-point} (o, {\tt
del punto cero}, $n=0$).  Como se menciona en muchos libros de
mec\'anica cu\'antica esta energ\'{\i}a est\'a relacionada con el
principio de incertidumbre --- no podemos medir exactamente la coordenada
y el momento lineal al mismo tiempo, lo que resulta en la existencia de
la energ\'{\i}a minima $E_0 \sim \hbar \omega/2$ del sistema
part\'{\i}cula -- campo potencial~\cite{Yariv}.  Desde mi punto de vista
este tipo de problemas demuestra la complicada estructura de vac\'{\i}o en
las teor\'{\i}as cu\'anticas e importancia del concepto de paridad.
Ademas, como todos los problemas cu\'anticos el sistema manifiesta el
principio de correspondencia, probablemente, el principio m\'as grande en
f\'{\i}sica y filosof\'{\i}a (vease, por ejemplo, ref.~\cite[\S
13]{Shiff}).\footnote{Estoy impresionado por su expresion por M.M. Nieto
en Memorias  del II Simposio ``Oscilador Arm\'onico": ``\ldots if you can
solve (or not solve) something classically the same is true quantum
mechanically, and {\it vice versa}".}

Entre las importantes aplicaciones del concepto  `interacci\'on
oscilador' (incluyendo los modelos para el problema de muchos cuerpos y los
modelos  con amortiguaci\'on) quisiera mensionar:

\begin{itemize}

\item
La representaci\'on de un n\'umero infinito de los osciladores del campo
y cuantizaci\'on secund\'aria en teor\'{\i}a de campos cuantizados
({\it e.g.}, el libro de N. N. Bogoliubov, 1973);

\item
Los espectros de las vibraciones de moleculas en el tratamiento
algebr\'aico ({\it e.g.}, los art\'{\i}culos de F. Iachello, A. Frank);

\item
En los modelos de c\'ascaras ({\it e.g.}, el libro de M. Mayer y J. Jensen,
1955) y modelos del movimiento colectivo, tales como el modelo
simpl\'ectico nuclear (D.  Rowe, 1977-85);

\item
La representaci\'on de los estados `squeeze' y los estados coherentes en
\'optica cu\'antica ({\it e.g.}, en las Mem\'orias de la ELAF'95, V. I.
Man'ko);

\item
Finalmente, el oscilador arm\'onico entr\'o en mec\'anica cu\'antica
relativista~\cite{Mosh}.

\end{itemize}

\section{Ecuaci\'on de Dirac y ecuaciones en
las representaciones m\'as altas}

De la mecanica cu\'antica relativista sabemos la relaci\'on entre
la energ\'{\i}a, la masa y el momento lineal:\footnote{El contenido de
esa Secci\'on tambien fue considerado desde diferentos puntos de
vista en el articulo antecedente de Dvoeglazov~\cite{DV-IC}. Ser\'{\i}a
\'util leer antes el art\'{\i}culo citado.}
\begin{equation} E^2 = {\bf
p}^{\,2} c^2 +m^2 c^4\quad. \label{rd} \end{equation}
La relaci\'on es
conectada \'{\i}ntimamente con las transformaciones de Lorentz de la
teor\'{\i}a de la relatividad~\cite[\S 2.3]{Ryder}.  Pero esa ecuaci\'on
no contiene informaci\'on acerca del espin, la variable adicional sin
fase~\cite{Wigner} y puede describir la evoluci\'on del campo escalar, o
part\'{\i}cula escalar, \'unicamente.  Despu\'es de la aplicaci\'on de la
transformaci\'on de Fourier obtenemos la ecuaci\'on de Klein-Gordon:
\begin{equation} \left ( {1\over c^2} {\partial^2 \over
\partial t^2} - \bbox{\nabla}^2 + {m^2 c^2 \over \hbar^2} \right ) \Psi
({\bf x}, t) = 0\quad.\label{kg}
\end{equation}
La funci\'on $\Psi ({\bf
x}, t)$ tiene una sola componente.  La energ\'{\i}a que corresponde a las
soluciones de la ec. (\ref{kg}) puede tener valores tanto negativos como
positivos. Aunque la densidad $\rho = {i\hbar \over 2mc^2} \left
(\Psi^\ast {\partial \Psi \over \partial t} - {\partial \Psi^\ast \over
\partial t} \Psi \right )$ satisface la ecuaci\'on de continuidad, hay
dificultades con su interpretaci\'on como densidad de
probabilidad, de acuerdo con la ecuaci\'on antecedente, $\rho$ puede ser
positiva o negativa (como la energ\'{\i}a) y no podemos ignorar las
soluciones con $E<0$ porque en este caso las soluciones con $E>0$ no
forman el sistema completo en sentido matem\'atico.  Por esas razones  en
los veintes y treintas buscaron las  ecuaciones para la funci\'on con
m\'as componentes y que sean lineales en la primer derivada respecto al
tiempo (como la ecuaci\'on de Schr\"odinger).  Aunque la ecuaci\'on lineal
de primer orden fue encontrada por P.  Dirac en 1928, sin cierta
reinterpretaci\'on ella ten\'{\i}a los mismos defectos. M\'as tarde W.
Pauli, V.  Weisskopf y M. Markov~\cite{Markov} argumentaron que $\rho$
se tiene que considerar como la densidad de carga y quitaron una de
las objeciones contra la teor\'{\i}a con las ecuaciones del segundo orden.

?` Como manejar
el grado de libertad de esp\'{\i}n en las ecuaciones de
segundo orden?  Sabemos  de la necesidad de introducirlo por los
experimentos como la separaci\'on de las lineas espectrosc\'opicas de la
part\'{\i}cula cargada en el campo magn\'etico, el efecto de Zeeman.  La
propuesta obvia para introducir el esp\'{\i}n es representar el operador
$(E^2/c^2) -{\bf p}^2$ como
\begin{equation}
\left ( {E^{(op)}\over c}
-{\bbox \sigma} \cdot {\bf p} \right ) \left ( {E^{(op)}\over c} + {\bbox
\sigma} \cdot {\bf p} \right ) =(mc)^2\label{eq-wdw}
\end{equation}
y  considerar la funci\'on como la  con dos componentes.  $E^{(op)} \equiv
i\hbar {\partial \over \partial t}$ y $\bbox{\sigma}$ son las matrices
de Pauli de la dimensi\'on $2\times 2$ .  La forma estandar de ellas es
\begin{eqnarray} \sigma_x=\pmatrix{0&1\cr 1&0\cr}\quad,\quad
\sigma_y=\pmatrix{0&-i\cr i&0\cr}\quad,\quad
\sigma_z=\pmatrix{1&0\cr 0&-1\cr}\quad. \label{mp}
\end{eqnarray}
Sin embargo, las diferentes representaciones de aquellos tambi\'en son
posibles~\cite[p.84]{Sakurai}.
La forma (\ref{eq-wdw}) es compatible con la relaci\'on  dispercional
relativista (\ref{rd}) gracias a las propiedades de las matrices de Pauli:
\begin{equation}
\bbox{\sigma}_i \bbox{\sigma}_j + \bbox{\sigma}_j \bbox{\sigma}_i =
2\delta_{ij} \quad,
\end{equation}
donde $\delta_{ij}$ es el simbolo de Kronecker.
En el espacio de coordenadas la ecuaci\'on se lee
\begin{equation}
\left ( i\hbar {\partial \over \partial x_0} + i\hbar \bbox{\sigma}\cdot
\bbox{\nabla} \right ) \left (i\hbar {\partial \over \partial x_0} -
i\hbar \bbox{\sigma}\cdot \bbox{\nabla} \right )\phi = (mc)^2
\phi\quad,\label{ww}
\end{equation}
donde $x_0 =ct$. Como mension\'o
Sakurai~\cite[p.91]{Sakurai} R. P. Feynman y L. M. Brown usaron esa
ecuaci\'on del segundo orden en derivadas en tiempo y fue perfectamente
v\'alido para un electr\'on. Pero, como sabemos de la teor\'{\i}a de las
ecuaciones diferenciales parciales, cuando tratamos de resolver una
ecuaci\'on de segundo orden (para predecir la conducta futura) necesitamos
especificar las condiciones iniciales para la funci\'on $\phi$ y su
primera derivada en el tiempo (la condici\'on adicional). En este punto hay
diferencia con lo que hizo Dirac en 1928 cuando propuse una ecuaci\'on
para electr\'on y positr\'on de primer orden en las derivadas. Me permito
recordar que el bispinor de Dirac $j=1/2$ es con cuatro componentes y el
problema puede ser resuelta si sabemos s\'olo la funci\'on en el instante
$t=0$.\footnote{Otras ecuaciones en la representaci\'on $(1/2,0)\oplus
(0,1/2)$ del grupo de Poincar\`e (para part\'{\i}culas del tipo Majorana,
estados auto/contr-auto conjugados de carga) fueron propuestas
recientamente por D. V. Ahluwalia, G. Ziino, A.  O. Barut y por
Dvoeglazov (1993-96), pero esas materias est\'an fuera de las metas del
presente art\'{\i}culo.} Concluyendo, podemos decir que el n\'umero
de  componentes independentes que tenemos que especificar para la
descripci\'on de una part\'{\i}cula cargada es cuatro, no importa si
usamos la ecuaci\'on de Dirac o la ecuaci\'on de Waerden (\ref{ww}).

Sin embargo, es posible reconstruir la ecuaci\'on de Dirac empezando desde
la forma (\ref{ww}). Para este objetivo vamos a definir \begin{equation}
\phi_{_R} \equiv {1\over mc} \left (i\hbar {\partial \over \partial x_0} -
i\hbar \bbox{\sigma}\cdot \bbox{\nabla} \right ) \phi\quad,\quad \phi_{_L}
\equiv \phi\quad.
\end{equation}
Entonces tenemos la equivalencia entre  la ecuaci\'on (\ref{ww}) y
el conjunto
\begin{mathletters}
\begin{eqnarray}
\label{d1} \left [i\hbar (\partial/\partial x_0) - i\hbar
\bbox{\sigma}\cdot \bbox{\nabla} \right ] \phi_{_L} &=& mc \phi_{_R}\quad,\\
\label{d2} \left [i\hbar (\partial/\partial x_0) + i\hbar
\bbox{\sigma}\cdot \bbox{\nabla} \right ] \phi_{_R} &=& mc \phi_{_L}\quad.
\end{eqnarray} \end{mathletters}
Tomando la suma y la diferencia de las
ecuaciones (\ref{d1},\ref{d2}) llegamos a la famosa ecuaci\'on
presentada por Dirac ($\psi = (\phi_{_R} +\phi_{_L})/\sqrt{2}$,\,\, $\chi
= (\phi_{_R} -\phi_{_L})/\sqrt{2}$):  \begin{equation} \pmatrix{i\hbar
(\partial /\partial x_0) & i\hbar \bbox{\sigma}\cdot \bbox{\nabla}\cr
-i\hbar \bbox{\sigma}\cdot \bbox{\nabla}& -i\hbar (\partial
/\partial x_0)\cr} \pmatrix{\psi \cr \chi} = mc \pmatrix{\psi \cr
\chi}\quad,
\end{equation}
o bien,
\begin{equation} \left [ i\gamma^\mu
\partial_\mu - m \right ] \Psi (x^\mu) = 0\quad,\quad \hbar=c=1\quad,
\label{ed}
\end{equation}
lo que coincide con~\cite[ec.(10)]{DV-IC}, \,   la ecuaci\'on para los
fermiones -- part\'{\i}culas con el espin $j=1/2$.
Las matrices de Dirac tienen la siguiente forma en esta representaci\'on
que se llama la representaci\'on estandar (o bien,
can\'onica):\footnote{Existe el  n\'umero infinito de las representaciones
de las matrices $\gamma$. Por eso se dice que esas matrices se
definen con la precisi\'on de la transformaci\'on unitaria.}
\begin{equation} \gamma^0 =\pmatrix{\openone & 0\cr 0
&-\openone\cr}\quad,\quad \bbox{\gamma}^i =\pmatrix{0&\bbox{\sigma}^i\cr
\bbox{-\sigma}^i &0\cr}\quad, \label{md} \end{equation} donde $\openone$ y
$0$ se deben entender como matrices de $2\times 2$.  La
ecuaci\'on de Klein-Gordon pudiera sido reconstruida de la ecuaci\'on
(\ref{ed}) despu\'es de multiplicaci\'on por el operador $\left [
i\gamma^\mu \partial_\mu + m\right ]$ que es otra `raiz  cuadrada' de la
ecuaci\'on (\ref{kg}). Este es otro
camino para deducir la ecuaci\'on de
Dirac\footnote{V\'ease Dvoeglazov~\cite{DV-IC} para otros dos.}, propuesto
por B. L. van der Waerden y J. J.  Sakurai~\cite{Sakurai}.

Como ya sabemos la funci\'on de Dirac tiene cuatro componentes complejos.
Pero, es posible saber la dimensi\'on de las matrices de algebra de Dirac
(que entran en la ecuaci\'on (\ref{ed})) en base de la simple deducci\'on
matem\'atica.  De la definici\'on  (\ref{md}) podemos concluir que
cuatro matrices $\gamma^\mu$  (o, bi\'en, $\alpha^\mu$, v\'ease la forma
hamiltoniana~\cite[ec.(8)]{DV-IC}) son anticomutados,
\begin{equation}
\gamma^\mu \gamma^\nu
+\gamma^\nu \gamma^\mu = 2g^{\mu\nu}\label{rc}
\end{equation}
con $g^{\mu\nu} = diag\,(1,-1,-1,-1)$
es el tensor de m\'etrica en el espacio de Minkowski.
Si $i\neq j$ tenemos
$\gamma_i \gamma_j +\gamma_j \gamma_i =0$.  Entonces, de acuerdo con las
reglas del \'algebra lineal:
\begin{equation} Det (\gamma_i \gamma_j) = Det
(-\gamma_j \gamma_i) = (-1)^d \, Det (\gamma_i \gamma_j )\quad,
\end{equation}
que nos da informaci\'on que la dimesi\'on tiene que ser
{\bf par}.  En el caso $d=2$ se tienen s\'olo tres matrices que
anticonmutan una con otra, son matrices de Pauli (\ref{mp}) que forman el
sistema completo en sentido matem\'atico.  Pero necesitamos cuatro
matrices $\gamma^1, \gamma^2, \gamma^3$ y $\gamma^0$. Conclu\'{\i}mos que
la dimensi\'on tiene que ser mayor o igual a cuatro $d \geq 4$.
La representaci\'on m\'as simple es $d=4$, v\'ease (\ref{md}).

En otras representaciones del grupo de Poincar\`e tambi\'en se pueden
proponer ecuaciones del primer orden, como hizieron de
Broglie, Duffin, Kemmer y Bhabha (por
ejemplo,~\cite{Fish}).\footnote{V\'ease acerca de las relaciones entre las
ecuaciones de primer y de segundo orden para las part\'{\i}culas con el
esp\'{\i}n $j=1$ en ref.~\cite[Secci\'on \# 4]{DV-IC}.} Ellos tienen la
forma de la ecuaci\'on de Dirac pero en los casos del esp\'{\i}n alto la
funci\'on del campo ya no es la funci\'on con cuatro componentes y las
matrices no satisfacen la relaci\'on de anticonmutaci\'on (\ref{rc}).
Pero en cada representaci\'on existen relaciones m\'as complicadas entre
esas matrices. Para el algebra de Duffin, Kemmer y Petiau, la
representaci\'on $(1,0)\oplus (0,1)\oplus (1/2,1/2)\oplus (1/2,1/2) \oplus
(0,0)\oplus (0,0)$, ellas se denotan como matrices $\beta$ (en lugar de
matrices $\gamma$) y satisfacen
\begin{equation} \beta_\mu \beta_\nu
\beta_\lambda + \beta_\lambda \beta_\nu \beta_\mu =\beta_\mu
g_{\nu\lambda} +\beta_\lambda g_{\mu\nu}\quad.
\end{equation}
Adem\'as, se
puede presentar la ecuaci\'on para las part\'{\i}culas escalares (el campo
escalar) en la forma de las derivadas de primer orden.
Introduciendo para la funci\'on de Klein-Gordon la siguiente notaci\'on
\begin{equation} \psi_0 \equiv {\partial \psi \over \partial t}\quad,\quad
\psi_i \equiv {\partial \psi \over \partial x^i}\quad,\quad
\psi_4 \equiv m\Psi
\end{equation}
el 'vector' con cinco componentes  satisface la ecuaci\'on
de primer orden, que ponen en la forma hamiltoniana~\cite{It}:
\begin{equation}
i{\partial \psi \over \partial t} = \left ({1\over i} \bbox{\alpha}
\cdot \bbox{\nabla} +m\beta \right) \psi\quad.
\end{equation}
La ecuaci\'on de Klein-Gordon se presenta tambi\'en en la
forma del conjunto de dos ecuaciones~\cite{Fish,Kos}
\begin{equation}
{\partial \Psi \over \partial x^\alpha}
=\kappa \Xi_\alpha\quad,\quad {\partial \Xi^\alpha \over \partial
x^\alpha} = -\kappa \Psi\quad, \label{kg10}
\end{equation}
con $\kappa \equiv mc/\hbar$,
que toma la forma de matrices siguiente:
\begin{equation} i{\partial \over \partial t}
\pmatrix{\phi\cr\chi_1\cr\chi_2\cr\chi_3\cr}
= \left [ \pmatrix{0&p_1&p_2&p_3\cr
p_1&0&0&0\cr
p_2&0&0&0\cr
p_3&0&0&0\cr} +m \pmatrix{1&0&0&0\cr
0&-1&0&0\cr
0&0&-1&0\cr
0&0&0&-1\cr}\right ]\pmatrix{\phi\cr\chi_1\cr\chi_2\cr\chi_3\cr}\quad,
\label{kg1}
\end{equation}
donde
\begin{eqnarray}
\cases{\phi = i\partial_t \Psi +m\Psi &\cr
\chi_i = -i\bbox{\nabla}_i \Psi = {\bf p}_i \Psi &\cr}\quad.
\end{eqnarray}

Finalmente, gracias a Dowker~\cite{Dowker} sabemos que una part\'{\i}cula
de cualquier esp\'{\i}n puede ser descrita por el sistema de
ecuaciones de primer orden:
\begin{mathletters} \begin{eqnarray}
\alpha^\mu \partial_\mu \Phi &=& m\Upsilon\quad,\label{dow1}\\
\overline{\alpha}^\mu \partial_\mu \Upsilon &=& -m\Phi \label{dow2} \quad.
\end{eqnarray} \end{mathletters}
$\Phi$ se transforma de acuerdo con la representaci\'on
$(j,0)\oplus (j-1,0)$ y $\Upsilon$ de acuerdo con
$(j-1/2,1/2)$. Las matrices $\alpha^\mu$ en este caso generalizado tienen
dimensi\'on $4j \times 4j$.

Muchas caracter\'{\i}sticas de las part\'{\i}culas pueden ser obtenidas
por el an\'alisis de la teor\'{\i}a de campos libres, por la ecuaciones
de la mec\'anica cu\'antica relativista que presentamos en esa Secci\'on.
Lo importante es prestar atenci\'on al formalismo matem\'atico
del grupo de Poincar\`e, desarollado basicamente por Wigner. Como
dir\'{\i}a el profesor A. Barut esa materia es muy viva hasta ahora. Pero,
la f\'{\i}sica siempre tiene muchos caminos: gracias al desarollo de los
aceleradores de altas energ\'{\i}as la tarea de los f\'{\i}sicos en
los \'ultimos cincuenta a\~nos era explicar los procesos con el cambio del
n\'umero de part\'{\i}culas,  para este objetivo era necessario desarollar
la met\'odica de calculaciones, tales como la met\'odica de diagramos de
Feynman, la teor\'{\i}a de la matriz $S$, modelos potenciales etc.
Gran parte de esos c\'alculos se basan en el concepto de la
interacci\'on, principalmente el concepto de la interacci\'on minimal,
$\partial_\mu \rightarrow \partial_\mu -ieA_\mu$, donde $A_\mu$ es el
potencial 4-vector.  Pero, matem\'aticamente, es posible introducir otros
tipos de interacci\'on como lo hizo el  doctor Moshinsky.

\section{Oscilador de Dirac de Moshinsky}

El concepto del oscilador arm\'onico relativista fue propuesto
por primera vez hace mucho tiempo~\cite{Ito} pero fue olvidado
y redescubierto en 1989 por el doctor Marcos Moshinsky~\cite{Mosh}.
En el caso del problema de un cuerpo \'el caracteriza por la
siguiente substituci\'on de interacci\'on {\bf no-minimal}
en la ecuaci\'on de Dirac:
\begin{equation}
{\bf p}
\rightarrow {\bf p}-im\omega {\bf r} \beta\quad,
\end{equation}
donde $m$ es la masa del fermi\'on, $\omega$ es la frequencia del
oscilador, ${\bf r}$ es la coordenada 3-dimensional, y $\beta \equiv
\gamma^0$ es una de matrices de algebra de Dirac (que tambi\'en es la
matriz del operador de paridad). Existen muy pocos problemas
de interacci\'on de una part\'{\i}cula, tales como a) potencial de
Coulomb; b) campo magn\'etico uniforme; c) la onda electromagn\'etica
plana, que se puede resolver en forma exacta en mec\'anica cu\'antica.  El
oscilador de Dirac de Moshinsky es un de ellos y un poco parecida al
problema b).\footnote{En el problema de la part\'{\i}cula en un campo
magn\'etico uniforme tenemos los terminos adicionales $\sim ({\bf A}\cdot
{\bf p})$. Es tarea para el lector comprobar los c\'alculos
en~\cite[p.67]{It} y compararlos con el problema que consideramos en el
texto de este art\'{\i}culo.} Las soluciones del conjunto para
2-espinores~\cite{Mosh}
\begin{mathletters}
\begin{eqnarray}
\label{do1}
(E-mc^2) \psi &=& c \bbox{\sigma} \cdot ({\bf p} +im\omega {\bf r})
\chi\quad,\\
\label{do2} (E+mc^2) \chi &=& c \bbox{\sigma} \cdot ({\bf
p} -im\omega {\bf r}) \psi
\end{eqnarray} \end{mathletters}
se han dados  por el {\tt ket}
\begin{equation}
\vert N (l {1\over 2}) jm> =
\sum_{\mu\sigma} <l\mu, {1\over 2} \sigma\vert  jm> R_{_{Nl}} (r)
Y_{_{lm}} (\theta,\phi) \chi_\sigma\quad.
\end{equation}
El espectro de energ\'{\i}a es entonces
\begin{equation}
(mc^2)^{-1} (E_{_{Nlj}}^2 - m^2 c^4) =
\cases{\hbar \omega \left [ 2(N-j)+1 \right ]\quad,\quad
\mbox{if}\,\,\,\,\, l=j-{1\over 2}&\cr \hbar \omega \left [ 2(N+j)+3
\right ]\quad,\quad  \mbox{if}\,\,\,\,\, l=j+{1\over 2}&\quad.\cr}
\end{equation}
Fue demostrado en los art\'{\i}culos~\cite{Moreno1} que la ecuaci\'ones
(\ref{do1},\ref{do2}) pueden ponerse en la forma covariante:
\begin{equation}
\left ( \hat p - mc  +\kappa {e\over 4m} \sigma^{\mu\nu}
F_{\mu\nu} \right ) \Psi = 0\quad, \quad\kappa =2m^2\omega/e\quad,
\label{fc}
\end{equation}
que  significa que la interacci\'on `oscilador' en el problema de un
cuerpo es esencialmente la interacci\'on tensorial con el campo el\'ectrico
(!` {\bf no} con el vector potencial!). Aunque en esa consideraci\'on
$F^{\mu\nu} = u_\mu x_\nu -u_\nu x_\mu$ con $u_\mu  = (1, {\bf 0})$ {\it
i.e.} es dependiente del sistema de referencia, no es dif\'{\i}cil
aplicar las transformaciones de Lorentz ({\tt boost} y rotaciones) para
reconstruir todos los resultados para los observables  f\'{\i}sicos de
cualquier sistema inercial. Dos notas que pudiera ser \'utiles para las
investigaciones futuras: 1) El vector $u^\mu$ pudiera ser utilizado para
la definici\'on de la parte transversal de $x^\mu$, a saber $x^\mu_{\perp}
\equiv x^\mu + (x^\nu u_\nu) u^\mu$; 2) La necesidad de la interacci\'on
tensorial ya fue aprobada en base del analisis de los datos experimentales
de los decaimientos de $\pi^-$ y $K^+$ mesones (V. N.  Bolotov {\it et
al.}, S. A.  Akimenko {\it et al.}, 1990-96).

Como se ha demostrado en  unos art\'{\i}culos {\it
e.g.}~\cite{Moreno2}, ese tipo de interacci\'on preserva la {\tt
supersimetr\'{i}a} de Dirac, el caso particular de {\tt
supersimetr\'{\i}a}.  Generalmente, el concepto de {\tt
supersimetr\'{\i}a} se define en el sentido de teor\'{\i}a de grupos como
una algebra:
\begin{mathletters} \begin{eqnarray}
\left \{ \hat Q\,\,, \hat
Q\,\,\right \}_+ &=&  \left \{ \hat Q^{\,\dagger}, \hat Q^{\,\dagger}\right
\}_+ =0\quad,\label{ss1}\\
\left [ \hat Q, \hat {\cal H}\right ]_- &=& \left [\hat
Q^{\,\dagger}, \hat {\cal H}\right ]_- =0\quad, \label{ss2}\\
\left \{
\hat Q, \hat Q^{\,\dagger} \right \}_+ &=& \hat{\cal H}\quad.\label{ss3}
\end{eqnarray} \end{mathletters}
$\hat Q^{\,\dagger}$ y $\hat Q$ se llama
{\tt supercargas}.  En el caso de la mec\'anica cu\'antica relativista de
particulas cargadas con espin $j=1/2$, el hamiltoniano se ha dado
por~\cite{Moreno2}
\begin{equation} \hat {\cal H} = Q +Q^{\,\dagger}
+\lambda\quad, \label{dh} \end{equation}
$\lambda$ es hermitiana. Entonces, si
\begin{equation}
\left \{ Q,\,\lambda \right \}
=\left \{ Q^{\,\dagger},\,\lambda \right \} =0 = Q^{\,2} =
Q^{\,\dagger^{\,2}}\label{dss1}
\end{equation}
tenemos
\begin{equation}
\left \{ Q, \,Q^{\,\dagger} \right \} = \hat{\cal H}^2 - \lambda^2
\quad.\label{dss2}
\end{equation}
Por ejemplo, si
\begin{equation}
Q =\pmatrix{0&0\cr \bbox{\sigma}\cdot ({\bf p} -im\omega {\bf
r})&0\cr}\quad,
\quad \mbox{y}\quad,
\quad Q^{\,\dagger} =
\pmatrix{0&\bbox{\sigma}\cdot ({\bf p} +im\omega{\bf r})\cr 0&0\cr}
\end{equation}
todas las condiciones (\ref{dh},\ref{dss1},\ref{dss2}) se satisfacen y de
(\ref{dss2}) obtenemos la ecuaci\'on del oscilador de Dirac.
Otros tipos del oscilador arm\'onico relativista para el problema de un
cuerpo han sido propuestas en~\cite{DV-HJ}, es interesante que \'estos
est\'an relacionados con la interacci\'on con la carga quiral o con
coplamiento pseudoescalar, $m \rightarrow m \left [1+(w/c) r\gamma_5
\right ]$.

El profesor Moshinsky dijo  en muchos seminarios que sus objetivos al
inventar ese tipo de interacci\'on eran aplicarlo al problema cu\'antico
relativista de muchos cuerpos.  Aunque existe el formalismo de Bethe y
Salpeter~\cite{BS} y los m\'etodos para manejar este
formalismo~\cite{DV-PPN} con el tiempo relativo,\footnote{Recuerda las
palabras de Eddington: ``Un electron ayer y un proton hoy no forman el
atomo de hidrogeno".}  fueron
aprobados en base a la comparaci\'on de los resultados te\'oricos y del
experimento, no todos f\'{\i}sicos quieren usarlo, principalmente, por su
complejidad. Unos problemas de descripci\'on alternativa han sido
considerados~\cite{Barut,Mosh2,Mosh3,Mosh4} desde diferentes puntos de
vista.  Las interacci\'ones del tipo `oscilador' han sido propuestas para
la ``ecuaci\'on de Dirac" para dos cuerpos. En este caso podemos
considerar la ecuaci\'on \begin{eqnarray} \left [ (\bbox{\alpha}_1 -
\bbox{\alpha}_2)\cdot ({\bf p} -i{m\omega \over 2} {\bf r} {\cal B}) +mc
(\beta_1 +\beta_2) \right ] \psi ={E\over c}\psi\label{dodc} \quad.
\end{eqnarray}
Los \'{\i}ndices $1$ y $2$ indican el
espacio de representaci\'on de primera o segunda part\'{\i}cula. En lugar
de la matriz ${\cal B}$ pueden ser substituidos $B=\beta_1 \beta_2$ o
$B\Gamma_5=\beta_1 \beta_2 \gamma^5_1 \gamma^5_2$. Entonces, tenemos dos
tipos de oscilador para dos cuerpos.  Los espectros  son
parecidos~\cite{Mosh5}. Adem\'as, ambos dan los valores propios de
energ\'{\i}a $E=0$, {\tt  relativistic cockroach nest
(RCN)}~\cite{Mosh2,Mosh5}, como les nombr\'o el doctor
Moshinsky.\footnote{Pienso que al problema del RCN se requiere m\'as
atenci\'on. \'El puede ser relacionado con las soluciones con $E=0$ que
eran descubiertas en otros sitemas f\'{\i}sicos (por ejemplo, en las
conocidas ecuaciones para el campo tensorial antisim\'etrico de segundo
rango, as\'{\i} como en las ecuaciones de primer orden para $j=3/2$ y
$j=2$, el \'ultimo es el campo gravitacional en la $(2,0)\oplus (0,2)$
representaci\'on) por D.  V.  Ahluwalia, A.  E.  Chubykalo, M.  W.  Evans
y J.-P.  Vigier, y por V. V. Dvoeglazov.  Pero esa materia tiene que ser
discutida en un art\'{\i}culo aparte.}

Las contribuciones de otros grupos cient\'{\i}ficos
ser\'an consideradas en la siguiente Secci\'on.

\section{Interacci\'on `oscilador' para
las part\'{\i}culas con espin alto.}

Las ecuaciones con la interacci\'on `oscilador' relativista han sido
tratados en los
art\'{\i}culos~\cite{Deb,Bruce,Ned,DV-NASA,DV-NC1,DV-NC2,DV-RMF1,DV-RMF2}
desde diversos puntos de vista. Ellos son para los espines diferentes del
espin $j=1/2$.

El operador de coordenada y el operador de
momento~\cite{Bruce} lineal han sido escogidos como $n\times n$ matrices:
$\widehat {\bf Q} = \hat \eta {\bf q}$ y $\widehat {\bf P} = \hat \eta
{\bf p}$.  La interacci\'on introducida en la ecuaci\'on de
Klein-Gordon fue entonces
$\widehat{\bf P} \rightarrow \widehat {\bf P} -im\hat
\gamma \hat \Omega \cdot \widehat{\bf Q}$. Las matrices satisfacen las
condiciones \begin{equation} \hat \eta^2 =\openone\quad,\quad \hat
\gamma^2 = \openone\quad,\quad \mbox{y}\quad \left \{ \hat \gamma, \hat
\eta \right \}_+ =0\quad.
\end{equation}
$\Omega$ es la matriz de las
frecuencias , de dimensi\'on $3\times 3$.  Como resultado tenemos el
oscilador anisotr\'opico en tres dimensiones:  \begin{equation}
-{\partial^2 \over \partial t^2} \Psi ({\bf q}, t) = \left ( {\bf p}^{\,2}
+m^2 {\bf q}\cdot \hat \Omega^2 \cdot {\bf q} +m\hat \gamma \,\mbox{tr}
\Omega +m^2 \right ) \Psi ({\bf q},t)\quad, \end{equation} donde la forma
explicita de las matrices constituyentes es
\begin{equation} \hat \eta
=\pmatrix{0&1\cr 1&0\cr}\quad,\quad \hat \gamma =\pmatrix{-1&0\cr
0&1\cr}\quad.
\end{equation}
El espectro en el l\'{\i}mite no relativista llega a ser el espectro del
oscilador anisotr\'opico.

Pero las razones de introducir la forma matricial
en la ecuaci\'on de Klein-Gordon no han sido claros en este
art\'{\i}culo.  En otros trabajos~\cite{DV-NASA,DV-NC1} otro
formalismo para la desripci\'on de part\'{\i}cula con $j=0$ y $j=1$ es
presentado.  Como mencionamos, la ecuaci\'on de Klein-Gordon puede
ser presentada en la forma (\ref{kg10}).  Entonces, la interacci\'on
`oscilador' se introduce a la ecuaci\'on (\ref{kg1}) en la misma manera:
${\bf p} \rightarrow {\bf p} -im\omega \beta {\bf r}$ con $\beta$,
la matriz ante el t\'ermino de masa en (\ref{kg1}).  En caso $\omega_1
=\omega_2 =\omega_3 \equiv \omega$ la ecuaci\'on resultante coincide con
(10a) de ref.~\cite{Bruce}.

Las ecuaci\'ones para la interacci\'on `oscilador' de
part\'{\i}culas con $j=0$ y $j=1$
tambi\'en se han discutido
en~\cite{Deb,Ned,DV-RMF1,DV-RMF2}.  La ecuaci\'on hamiltoniana en el
formalismo de Duffin, Kemmer y Petiau toma la forma\footnote{Recuerda que
las matrices $\beta$ no anticonmutan, entonces, la reducci\'on de la
ecuaci\'on covariante a forma hamiltoniana es m\'as complicada. Adem\'as,
es necesario mencionar que en el proceso de deducci\'on de la forma
hamiltoniana~\cite{Kemmer} los autores hicieron un procedimiento
matem\'aticamente dudoso cuando se multiplic\'o la ecuaci\'on por la
matriz singular $\beta_0$. Sin embargo, depende del lector  si ser de
acuerdo con la discuci\'on en p. 110 de ref.~[32a].}
\begin{equation} i{\partial
\Phi \over \partial t} = \left ( {\bf B}\cdot {\bf p} +m\beta_0 \right )
\Phi\quad,\quad B_\mu = \left [\beta_0, \beta_\mu \right ]_- \quad,
\end{equation} donde $\Phi$ es la
funci\'on con 5 componentes en el caso $j=0$ y con 10 componentes en el
caso $j=1$.  La interaci\'on que introducen N. Debergh {\it et
al.}~\cite{Deb} es ${\bf p} \rightarrow {\bf p}-im\omega \eta_0 {\bf r}$,
\,\,$\eta_0 =2\beta_0^2 -1$. Como el resultado, en el limite no
relativista se encuentra el mismo t\'ermino ${1\over 2} m\omega^2 {\bf
r}^{\,2}$  que en el caso de la representaci\'on $(1/2,0)\oplus (0,1/2)$
anterior. Adem\'as, tambi\'en tenemos el coplamiento espin-orbita.
Este es el oscilador de Duffin y Kemmer. La misma
substituci\'on nominimal fue discutida por Nedjadi y Barrett~\cite{Ned}
pero fue introducida en la forma covariante de la ecuaci\'on de Duffin,
Kemmer y Petiau (v\'ease footnote \# 10).

En otro art\'{\i}culo se empes\'o desde el sistema de
ecuaciones de Bargmann-Wigner (ecs. (2,3) en ref.~\cite{DV-RMF1})
y se consider\'o la funci\'on de Bargmann-Wigner antisim\'etrica en los
indices espinoreales (ec. (4)en ref.~\cite{DV-RMF1}).  Los resultados de
ese art\'{\i}culo llegan a una conclusi\'on acerca de la existencia de los
estados doble degenerados en $N$, el numero cu\'antico principal, en el
l\'{\i}mite $\hbar \omega << mc^2$ excepto el nivel de base.

Los t\'erminos de interacci\'on en tres art\'{\i}culos citados
pueden ser presentados en forma covariante como los t\'erminos de
interacci\'on  de la forma $\kappa S^{\mu\nu} F_{\mu\nu}$, {\it cf.}
con~(\ref{fc}).  Vamos a componer una tabla
en que se comparan las formas de interacci\'on de varios art\'{\i}culos:\\

\medskip

\begin{tabular}{cc}
\hline
Referencia & El termino de interacci\'on\\
\hline
&\\
\cite{Deb}& $S^{\mu\nu} = -2i \left \{ \beta^\mu,\, B^\nu\right \}_+$\\
&\\
\cite{Ned}& $S^{\mu\nu}=\beta^\mu \eta^\nu$\\
&\\
\cite{DV-RMF1}& $S^{\mu\nu} = \beta^\mu
\beta^\nu - \beta^\nu \beta^\mu$ \\
&\\
\hline
\end{tabular}

Tambi\'en han sido considerados:

\begin{itemize}

\item
La interacci\'on `oscilador' en el formalismo de Sakata y
Taketani~\cite{Deb,DV-NASA,DV-NC1};

\item
El oscilador de Dirac en $(1+1)$ dimensi\'on~\cite{Dom,DV-NC1};

\item
El oscilador de Dirac en la forma con cuaterniones~\cite{DV-NASA};

\item
El oscilador para el sistema de ecuaciones de Dowker
(\ref{dow1},\ref{dow2}), lo que significa que este tipo de
interacci\'on puede ser introducido para cualquier espin~\cite{DV-NC2};

\item
El oscilador en el $2(2j+1)$ formalismo~\cite{DV-NASA};

\item
El oscilador de Dirac para el sistema de dos cuerpos
y su conexi\'on con el formalismo de Proca, y de
Bargmann y Wigner~\cite{DV-NASA,DV-RMF2}.

\end{itemize}

Finalmente, del conjunto de las ecuaciones de Crater y  van
Alstine~\cite{Crater} y Sazdjian~\cite{Sazdjian} para un problema
de dos cuerpos podemos deducir una ecuaci\'on muy parecida a la
ecuaci\'on de oscilador de Dirac para dos cuerpos con el t\'ermino
de potencial m\'as general~[22b]:
\begin{equation}
{\cal V}^{int} (r) = {1\over r} \frac{dV(r)/dr}{1- \left [ V(r) \right ]^2
} \, i \left ( \bbox{\alpha}_2 -\bbox{\alpha}_1 \right ) B \Gamma_5 {\bf
r} \quad.
\end{equation}
M. Moshinsky {\it et al.} escogieron $V(r) =
\tanh (\omega r^2/4)$ y obtuvieron el oscilador de Dirac con la
interacci\'on del segundo tipo $B \Gamma_5 {\bf r}$, v\'ease  (\ref{dodc}).
Nosotros queremos conectar esa formulaci\'on con las antecedentes y
proveer alguna base para escoger el potencial. Para este  objetivo
vamos recordar el art\'{\i}culo~\cite{Skachkov} donde el potencial en la
representaci\'on configuracional relativista
\begin{equation} V(r) = -g^2
\frac{\coth (rm\pi)}{4\pi r}
\label{ps}
\end{equation}
ha sido deducido en base del an\'alisis de la
serie principal y la serie complementaria del grupo de Lorentz y la
aplicaci\'on de las transformaciones de Shapiro en lugar de las
transformaciones de Fourier al potencial de Coulomb en el espacio del
momento lineal; la coordenada en el espacio configuracional se considera
en la forma m\'as general y puede ser imaginario $r \rightarrow i\rho$.
Esa forma de potencial encontr\'o algunas aplicaciones en los modelos
potenciales de cromodin\'amica cu\'antica y electrodin\'amica cu\'antica.
Entonces, el uso del potencial de Skachkov tiene razones definitivas. En
caso del uso del potencial (\ref{ps}) tenemos una  conducta
asintotica del termino ${\cal V}^{int}$ diferente en tres regiones. En
la regi\'on $r>> {1\over m\pi}$ \begin{eqnarray} {\cal V}^{int} (r) &\sim&
{1\over r \left [ r^2 -(g/4\pi)^2 \right ]} ( \bbox{\alpha}_1
-\bbox{\alpha}_2 ) B\Gamma_5 {\bf r}\approx\\ & & \approx \cases{(1/r^3) i
( \bbox{\alpha}_1 -\bbox{\alpha}_2 ) B\Gamma_5 {\bf r} \quad, \quad
\mbox{si}\quad r>>{1\over m\pi}\quad \mbox{y}\quad r> ({g\over 4\pi})^2
&\cr -(1/r) i(\bbox{\alpha}_1 -\bbox{\alpha}_2) B\Gamma_5 {\bf
r}\quad,\quad\mbox{si}\quad {1\over m\pi} << r < ({g\over
4\pi})^2\quad;&\cr}\nonumber
\end{eqnarray}
y en la regi\'on
$r<<{1\over \kappa}$
\begin{equation}
{\cal V}^{int} (r) \sim -im
(\bbox{\alpha}_1  -\bbox{\alpha}_2 ) B\Gamma_5
{\bf r}\quad,\quad\mbox{si}\quad r<<{1\over m\pi}\quad.
\end{equation}
Entonces, podemos ver que en la regi\'on de las distancias peque\~nas
tenemos precisamente la conducta del potencial del oscilador de Dirac.  En
la regi\'on de las distancias grandes tenemos la conducta del potencial de
inversos grados en $r$.

Como conclusi\'on: el concepto del `oscilador de Dirac' aunque
ha sido propuesto recientemente  se ha  desarrollado
mucho en los \'ultimos a\~nos pues nos permite describir bien diferentes
sistemas f\'{\i}sicos relativistas (incluyendo espectros de mesones y
bariones) desde un punto de vista diferente al punto de vista com\'un.
Esas ideas se publican en las revistas de mayor nivel internacional
como {\it Physical Review Letters}, {\it Physics Letters}, {\it Journal of
Physics} y {\it Nuovo Cimento}.  Por eso yo llamo a los
jovenes f\'{\i}sicos mexicanos aplicar sus talentos a esa area de
investigaciones.

Quiero indicar que en esa nota \'unicamente delineemos unos
rasgos de ese problema de la mec\'anica cu\'antica relativista
y no toquemos  muchas ideas (por ejemplo, las teor\'{\i}as
del electr\'on extendido con adicionales grados de libertad
intr\'{\i}nsecos~\cite{Dirac}) que pueden tener cierta relaci\'on con el
`oscilador de Dirac' pero que todav\'{\i}a no han sido desarrollados
suficientemente y no han sido aceptados por mucha gente.

\bigskip

Agradezco mucho al doctor D. Armando Contreras Solorio por su invitaci\'on
a trabajar en la Escuela de F\'{\i}sica de la Universidad Aut\'onoma de
Zacatecas y los doctores M. Moshinsky, Yu. F. Smirnov y A. Del Sol Mesa por
sus valiosas discusiones.  Reconozco la ayuda en la ortograf\'{\i}a
espa\~nola del Sr.  Jes\'us Alberto C\'azares.

\end{document}